\documentclass[preprint,showpacs,preprintnumbers,amsmath,amssymb, nofootinbib]{revtex4}
\usepackage{graphics,epsfig,subfigure}
\usepackage{color}
\usepackage[linktocpage]{hyperref}

\begin{document}
\renewcommand{\baselinestretch}{1.3}
\newcommand\beq{\begin{equation}}
\newcommand\eeq{\end{equation}}
\newcommand\beqn{\begin{eqnarray}}
\newcommand\eeqn{\end{eqnarray}}
\newcommand\nn{\nonumber}
\newcommand\fc{\frac}
\newcommand\lt{\left}
\newcommand\rt{\right}
\newcommand\pt{\partial}
\newcommand\tx{\text}
\newcommand\mc{\mathcal}

\allowdisplaybreaks

\title{Electrically charged black holes in gravity with a background Kalb-Ramond field}
\author{
Zheng-Qiao Duan,
Ju-Ying Zhao,
Ke Yang\footnote{keyang@swu.edu.cn, corresponding author}
}

\affiliation{School of Physical Science and Technology, Southwest University, Chongqing 400715, China}

\begin{abstract}

We derive the exact solutions for electrically charged black holes both in the absence and presence of a cosmological constant in the gravitational theory with Lorentz violation induced by a background Kalb-Ramond (KR) field. The corresponding thermodynamic properties are investigated. It is found that the standard first law of thermodynamics and the Smarr formula remain valid for the charged KR black holes. Nevertheless, the Lorentz-violating effect influences their ranges of local thermodynamic stability and the first- and second-order phase transition points. Furthermore, to examine the impact of Lorentz violation on the motion of test particles in the spacetime, we analyze the shadow and the innermost stable circular orbit (ISCO) of these black holes. Our results reveal that both the shadow and ISCO radii exhibit a high sensitivity to the Lorentz-violating parameter $\ell$, with a decrease observed as $\ell$ increases.

\end{abstract}







           


\maketitle



\section{Introduction}

Einstein's general relativity (GR) successfully explains a wide range of gravitational phenomena and is one of the two pillars of modern theoretical physics, along with quantum mechanics. It serves as the foundation for modern cosmology and provides us an accurate description of the evolution of our Universe. The GR extends the principles of special relativity to include the gravity, and maintains the local Lorentz symmetry at each point in spacetime manifold. Although current experiments and observations support Lorentz symmetry as a fundamental symmetry in the nature, some theoretical studies, particularly the string theory \cite{Kostelecky1989a}, loop quantum gravity \cite{Alfaro2002}, Horava-Lifshitz gravity \cite{Horava2009a}, noncommutative field theory \cite{Carroll2001}, suggest the possibility of Lorentz symmetry breaking (LSB) above some energy scales. These studies provide valuable insights into our understanding of the nature of spacetime and fundamental principles of physics.

A general theoretical framework for studying the LSB is the Standard-Model Extension, which extends the Standard Model of particle physics to include the GR and possible violations of Lorentz symmetry \cite{Kostelecky2004a}. Within this framework, a simple theory exhibiting LSB phenomenon is the bumblebee model \cite{Kostelecky1989a,Kostelecky1989,Kostelecky1989b,Bailey2006,Bluhm2008a}. The theory extends GR by incorporating a vector field $B_\mu$, known as the bumblebee field, which  non-minimally couples to gravity. When the bumblebee field acquires a nonzero vacuum expectation value (VEV), it serves as a fixed background field that spontaneously breaks the Lorentz symmetry. The bumblebee model has been intensively studied in various areas recently, including black hole physics \cite{Casana2018,Ovgun2018,Kanzi2019,Yang2019b,Cai2023,Oliveira2021,Maluf2021,Xu2023,Mai2023,Xu2023a,Liang2023,Ding2020a,Ding2021,Wang2022,Liu2019,Liu2023,Wang2022a,Ding2023,Chen2023,Gullu2022,Zhang2023,Lin2023,Ding2022,Jha2021,Ding2023a,Lessa2023,Carleo2022,Lambiase2023}, wormholes \cite{Ovgun2019}, cosmology \cite{Khodadi2023a,Khodadi2023}, and gravitational waves \cite{Liang2022,Amarilo2023}.

The KR field is a rank-two antisymmetric tensor arising naturally from the spectrum of bosonic string theory \cite{Kalb1974}, and its properties have been extensively studied in various contexts \cite{Kao1996,Kar2003,Fu2012,Nair2022}. An alternative approach to trigger the spontaneously LSB involves the consideration of a KR field $B_{\mu\nu}$ that non-minimally couples to gravity \cite{Altschul2010}. When the KR field acquires a nonzero VEV, it leads to the spontaneous breaking of Lorentz symmetry. A Schwarzschild-like solution was derived in the theory in Ref.~\cite{Lessa2020}. The corresponding particle motion, gravitational lensing and energy processes around the Schwarzschild-like black hole was investigated in Ref.~\cite{Atamurotov2022}. The quasinormal modes and bounding greybody factors of the GUP-corrected black holes was investigated in \cite{Baruah2023}. In Ref.~\cite{Kumar2020c}, a rotating KR black hole solution was obtained and the corresponding deflection of light and shadow cast by the black hole were investigated. In Refs.~\cite{Lessa2021,Maluf2022}, the traversable wormhole solutions were derived, and the corresponding gravitational lensing effect of the wormhole solutions was discussed in Refs.~\cite{Javed2022,Javed2023}. A consequence of LSB on the Bianchi type I cosmology was considered in Ref.~\cite{Maluf2022a}. In Ref.~\cite{Yang2023a}, a different Schwarzschild-like solution was derived in this theory and the Schwarzschild-(A)dS-like solution was obtained as well by relaxing the vacuum conditions. The corresponding shadows and quasinormal frequencies of these black holes were investigated in Ref.~\cite{Filho2023}.

Charged black holes exhibit unique properties due to the presence of electric charge, which can have significant effects on their geometry and thermodynamics. In this work, we are interested in constructing the electrically charged, static, and spherically symmetric black hole solutions in the gravitational theory with Lorentz violation triggered by a nonzero VEV background of the KR field. By examining the influence of Lorentz violation on their geometry, thermodynamics, and orbital motion of test particles, we aim to deepen our understanding of the unique properties exhibited by charged black holes in the Lorentz-violating gravity.

The layout of the paper is as follows: In Sect.~\ref{Model}, we introduce the theory and incorporate an interaction term between the KR field and the electromagnetic field. In Sect.~\ref{Solution}, we solve the theory to achieve analytical solutions for the electrically charged black holes both in the absence and presence of a cosmological constant. Further, several basic thermodynamic properties of the charged KR black holes are analyzed in Sect.~\ref{Thermodynamics}. Moreover, the impact of the Lorentz-violating effect on the motion of test particles near these black holes is investigated in Sect.~\ref{Orbits}. Finally, brief conclusions are present.

\section{Lorentz-violating gravity with a background KR field}\label{Model}

The KR field, denoted as $B_{\mu\nu}$, is a rank-two antisymmetric tensor field that obeys the condition $B_{\mu\nu}=-B_{\nu\mu}$. Its field strength is defined as a 3-form, denoted as $H_{\mu\nu\rho}\equiv \pt_{[\mu}B_{\nu\rho]}$. The field strength is invariant under the gauge invariance $B_{\mu\nu} \to B_{\mu\nu}+\pt_{[\mu}\Gamma_{\nu]}$. It is convenient to decompose the KR field to be $B_{\mu\nu}=\tilde E_{[\mu}v_{\nu]}+\epsilon_{\mu\nu\alpha\beta}v^\alpha \tilde B^\beta$ with $\tilde E_\mu v^\mu=\tilde B_\mu v^\mu=0$, where $v^\alpha$ is a timelike 4-vector \cite{Altschul2010,Lessa2020}. Therefore, the spacelike pseudo-vector fields $\tilde E_\mu$ and $\tilde B_\mu$ can be interpreted respectively as the pseudo-electric and pseudo-magnetic fields in analogy with Maxwell electrodynamics.

The action for the theory, which includes gravity non-minimally coupled to a self-interacting KR field, is given by \cite{Altschul2010,Lessa2020}
\beqn
S\!\!&\!=\!&\!\!\fc{1}{2}\!\int{}\!d^4x\sqrt{-g}\bigg[R-2\Lambda \!-\! \fc{1}{6}H^{\mu\nu\rho}H_{\mu\nu\rho}\!-\!V(B^{\mu\nu} B_{\mu\nu}\!\pm \!b^2) +\xi_2 B^{\rho\mu}B^{\nu}{}_\mu R_{\rho\nu}+\xi_3 B^{\mu\nu}B_{\mu\nu}R  \bigg] \nn\\
\!&\!+\!&\! \int{d^4x\sqrt{-g}\mathcal{L}_\tx{M}},
\label{Main_Action}
\eeqn
where $\Lambda$ represents the cosmological constant, $\xi_{2,3}$ are the non-minimal coupling constants between gravity and the KR field, and we have set $8\pi G=1$ for convenience. To achieve the charged solutions, we consider the matter Lagrangian $\mathcal{L}_\tx{M}$  to be the electromagnetic field, given by $\mathcal{L}_\tx{M} = -\frac{1}{2}F^{\mu\nu}F_{\mu\nu} + \mathcal{L}_\tx{int}$, where $F_{\mu\nu} = \partial_\mu A_\nu - \partial_\nu A_\mu$ represents the field strength of the electromagnetic field, and $\mathcal{L}_\text{int}$ represents the interaction between the electromagnetic field and the KR field.

The potential $V(B^{\mu\nu} B_{\mu\nu}\pm b^2)$ depends on $B^{\mu\nu}B_{\mu\nu}$ in order to maintain the theory's invariance under the observer local Lorentz transformation. As the cosmological constant $\Lambda$ is counted separately, the potential is set to zero at its minimum.  The minimum is determined by the condition $B^{\mu\nu} B_{\mu\nu}=\mp b^2$, with the sign $\pm$ chosen such that $b^2$ is a positive constant. Correspondingly, the KR field acquires a nonzero VEV, denoted as $\langle B_{\mu\nu} \rangle=b_{\mu\nu}$. Due to the non-minimal coupling of the KR field to gravity, the nonzero VEV background  $b_{\mu\nu}$ spontaneously breaks particle local Lorentz invariance.  In the vacuum configuration, the interaction term $\xi_3 B^{\mu\nu}B_{\mu\nu}R = \mp \xi_3 b^2 R$ in the action \eqref{Main_Action} can be absorbed into the Einstein-Hilbert terms by redefining variables.

Furthermore, we take the assumption that the only non-vanishing components of the vacuum configuration of the KR field are given by \cite{Lessa2020}
\beq
b_{10}=-b_{01}=\tilde E(r). \label{KR_VEV}
\eeq
Consequently, this configuration automatically vanishes the KR field strength, i.e., $H_{\lambda\mu\nu} = 0$.

In order to achieve the electrically charged black hole solutions, we consider an electrostatic vector potential $A_\mu=-\Phi(r)\delta^t_\mu$ in the usual manner. However, it is important to note that a consistent charged black hole solution cannot be supported solely by the electromagnetic field without the interaction term $\mathcal{L}_\text{int}$. To incorporate the interaction, one approach is to modify the KR field strength $H_{\mu\nu\rho}$ by adding a $U(1)$ electromagnetic Chern-Simons three-form, i.e., $\tilde H_{\mu\nu\rho}=H_{\mu\nu\rho}+A_{[\mu}F_{\nu\rho]}$ \cite{Majumdar1999}. However, for the vacuum KR configuration \eqref{KR_VEV} and the electrostatic vector potential, it is found that all the interactions in the modified kinetic term $\tilde H^{\mu\nu\rho}\tilde H_{\mu\nu\rho} = H^{\mu\nu\rho} H_{\mu\nu\rho} + 2H^{\mu\nu\rho}A_{[\mu}F_{\nu\rho]} + A^{[\mu}F^{\nu\rho]} A_{[\mu}F_{\nu\rho]}$ still vanish. Therefore, in order to introduce a nontrivial contribution to the spacetime dynamics, we consider the Lagrangian as the form                                                                                                                                                                                                                                                                                                                                                                                                                                                                                                                                                                                                                                                                                                                                                                                                                                                                                                                                                                                                                                                                                                                                                                                                                                                     
\beq
\mathcal{L}_\tx{M}=-\frac{1}{2}F^{\mu\nu}F_{\mu\nu}-\eta B^{\alpha\beta}B^{\gamma\rho}F_{\alpha\beta}F_{\gamma\rho},
\label{Interaction}
\eeq 
where $\eta$ is a coupling constant. When the KR field acquires a nonzero VEV, the Lagrangian induces LSB of the electromagnetic field, and allows for the existence of electrically charged black hole solutions.

The modified Einstein equations is obtained by varying the action \eqref{Main_Action} with respect to the metric $g^{\mu\nu}$, given by 
\beqn
R_{\mu \nu }-\frac{1}{2}g_{\mu \nu }R+\Lambda  g_{\mu \nu }= T^{\tx{M}}_{\mu\nu} + T^{\tx{KR}}_{\mu\nu},
\label{EoM1}
\eeqn
where $T^{\tx{M}}_{\mu\nu}$ is the energy-momentum tensor of the electromagnetic field, derived as
\beqn
T^{\tx{M}}_{\mu\nu}&=& 2 F_{\mu \alpha } F_{\nu }{}^{\alpha }- \fc{1}{2} g_{\mu \nu }F^{\alpha \beta } F_{\alpha \beta } +\eta \lt(8  B^{\alpha \beta } B_{\nu }{}^{\gamma } F_{\alpha \beta }  F_{\mu \gamma } - g_{\mu \nu } B^{\alpha \beta } B^{\gamma \rho } F_{\alpha \beta } F_{\gamma \rho } \rt),
\eeqn
 and $T^{\tx{KR}}_{\mu\nu}$ is the effective energy-momentum tensor of the KR field, given by
\beqn
T^{\tx{KR}}_{\mu\nu} \!&\!=\!&\!\frac{1}{2} H_{\mu \alpha \beta } H_{\nu }^{\alpha \beta }-\frac{1}{12} g_{\mu \nu } H^{\alpha \beta \rho } H_{\alpha \beta \rho }+2V' B_{\alpha\mu}B^{\alpha}{}_\nu -g_{\mu\nu}V \nn\\
\!&\!+ \!&\!\xi_2 \bigg[\frac{1}{2} g_{\mu \nu } B^{\alpha \gamma } B^{\beta }{}_{\gamma }R_{\alpha \beta } - B^{\alpha }{}_{\mu } B^{\beta }{}_{\nu }R_{\alpha \beta } -B^{\alpha \beta } B_{\nu \beta } R_{\mu \alpha }
\nn\\
\!&\!-\!&\!B^{\alpha \beta } B_{\mu \beta } R_{\nu \alpha }\!+\!\frac{1}{2} \nabla _{\alpha }\nabla _{\mu }\left(B^{\alpha \beta } B_{\nu \beta }\right)+\frac{1}{2} \nabla _{\alpha }\nabla _{\nu }\left(B^{\alpha \beta } B_{\mu \beta }\right)
\nn\\
\!&\!-\!&\!\frac{1}{2}\nabla ^{\alpha }\nabla _{\alpha }\left(B_{\mu }{}^{\gamma }B_{\nu \gamma } \right)-\frac{1}{2} g_{\mu \nu } \nabla _{\alpha }\nabla _{\beta }\left(B^{\alpha \gamma } B^{\beta }{}_{\gamma }\right)\bigg].
\eeqn
Here, the prime represents the derivative with respect to the argument of the corresponding functions. Note that the total energy-momentum tensor $T^{\tx{KR}}_{\mu\nu}+T^{\tx{M}}_{\mu\nu}$ is conserved due to the Bianchi identities.

The equation of motion for the KR field is obtained by taking the variation of the action \eqref{Main_Action} with respect to $B^{\mu\nu}$. It can be expressed as
\beq
\nabla ^{\alpha }H_{\alpha \mu \nu }\!+\!3\xi R_{\alpha [\mu }B^{\alpha }{}_{\nu ]}\!-\!6 V' B_{\mu \nu } \!-\!12 \eta  B^{\alpha \beta } F_{\alpha \beta } F_{\mu \nu }\!=\!0.
\label{EOM_KR}
\eeq
Furthermore, the modified Maxwell equation is derived by varying the action \eqref{Main_Action} with respect to the vector potential $A^{\mu}$, yielding
\beq
\nabla ^{\nu }\lt(F_{\mu \nu } + 2\eta  B_{\mu \nu }B^{\alpha \beta }  F_{\alpha \beta }\rt)=0.
\label{Maxwell_EQ}
\eeq
It reduces to the standard Maxwell equation when the coupling constant $\eta$ is set to zero.

\section{Electrically charged black hole solutions}\label{Solution}

We consider the metric ansatz for a static and spherically symmetric spacetime given by
\beq
ds^2=-F(r)dt^2+G(r)dr^2+r^2 d\theta^2+r^2 \sin^2\theta d\phi^2.
\label{Metric}
\eeq
With the ansatz, the function $\tilde E(r)$ \eqref{KR_VEV} can be further expressed as $\tilde E(r)=|b|\sqrt{\frac{F(r)G(r)}{2}}$. Consequently, the vacuum configuration of the KR field satisfies the constant norm condition $b^{\mu\nu}b_{\mu\nu}=-b^2$. 

Under the vacuum configuration, it is advantageous to reformulate the modified Einstein equation \eqref{EoM1} as
\beqn
R_{\mu\nu}\!&\!=\!&\!T^{\tx{M}}_{\mu\nu}-\fc{1}{2}g_{\mu\nu}T^{\tx{M}}+\Lambda  g_{\mu \nu }+V' \left(2 b_{\mu \alpha } b_{\nu }{}^{\alpha }+b^2 g_{\mu \nu }\right) \nn\\
\!&\!+\!&\! \xi_2 \bigg[g_{\mu \nu } b^{\alpha \gamma }  b^{\beta }{}_{\gamma }R_{\alpha \beta }- b^{\alpha }{}_{\mu } b^{\beta }{}_{\nu }R_{\alpha \beta } -b^{\alpha \beta } b_{\mu \beta } R_{\nu \alpha }-b^{\alpha \beta } b_{\nu \beta } R_{\mu \alpha }\nn\\
\!&\!+\!&\! \frac{1}{2} \nabla _{\alpha }\nabla _{\mu }\left(b^{\alpha \beta } b_{\nu \beta }\right)\!+\!\frac{1}{2} \nabla _{\alpha }\nabla _{\nu }\lt(b^{\alpha \beta } b_{\mu \beta }\rt)-\frac{1}{2}\nabla ^{\alpha }\nabla _{\alpha }\lt(b_{\mu }{}^{\gamma }b_{\nu \gamma } \rt)\bigg],
\label{EoM2}
\eeqn
where $T^{\tx{M}} \equiv g^{\alpha\beta}T^{\tx{M}}_{\alpha\beta}$.

Further, with the ansatzes of the metric  \eqref{Metric} and the electrostatic field, the field equations \eqref{EoM2} can be written explicitly as
\begin{subequations}
\beqn\label{EOM_Exp}
\frac{2 F''}{F}-\fc{F'}{F}\frac{ G'}{ G}-\frac{F'^2}{F^2}+\fc{4}{r}\frac{F'}{F}+\frac{4 \Lambda G }{1-\ell}
-\frac{4 \left(1 - 2\eta b^2  \right)\Phi'^2}{(1-\ell)F}=0,\label{EoM_1}\\
\frac{2 F''}{F}-\fc{F'}{F}\frac{ G'}{ G}-\frac{F'^2}{F^2}-\fc{4}{r}\frac{G'}{G}+\frac{4 \Lambda G }{1-\ell}
-\frac{4 \left(1 - 2\eta b^2  \right)\Phi'^2}{(1-\ell)F}=0,\label{EoM_2}\\
\frac{2 F''}{F}-\frac{F' G'}{F G}-\frac{F'^2}{F^2}+\frac{1+\ell}{\ell  r}\left(\frac{F'}{F}-\frac{G'}{G}\right)+\frac{2 (1-\ell )}{\ell  r^2}
-\left(1-\Lambda r^2-b^2 r^2 V'\right)\frac{2 G }{\ell r^2}\qquad\nn\\ -\frac{ 2\left(1 - 6\eta b^2  \right)\Phi'^2}{\ell F}=0,\label{EoM_3}
\eeqn
\end{subequations}
where $\ell\equiv \xi_2b^2/2$, referred to as the Lorentz-violating parameter, measures the magnitude of Lorentz-breaking effects. In addition, the field equation for the KR field \eqref{EOM_KR} and the modified Maxwell equation \eqref{Maxwell_EQ} are written explicitly as
\beqn
\fc{2F''}{F}-\fc{F'^2}{F^2}+\fc{2}{r}\lt(\fc{F'}{ F}-\fc{G'}{G}\rt)-\fc{F'}{F}\fc{G'}{G}  +\fc{2 b^2 V' G}{l}-\fc{8\eta b^2 \Phi'^2}{\ell F}=0,\label{EOM_KR_2}\\
\lt(1-2 \eta b^2 \rt) \lt[\Phi''+\frac{\Phi'}{2} \lt(\fc{4}{r}-\fc{F'}{F}-\frac{G'}{G}\rt)\rt]=0.
\label{Maxwell_EQ_2}
\eeqn

\subsection{Case: $\Lambda = 0$}

When the cosmological constant is absent, we take the assumption that $V'=0$, which corresponds to the case where the VEV is located at the local minimum of the potential. For instance, it can be simply realized by a potential of quadratic form, $V=\fc{1}{2}\lambda X^2$, with $X \equiv B^{\mu\nu} B_{\mu\nu}+ b^2$ and $\lambda$ a coupling constant \cite{Bluhm2008}.

By subtracting Eq.~\eqref{EoM_2} from Eq.~\eqref{EoM_1}, we have the following relation 
\beqn
\frac{F'}{F}=-\frac{G'}{G}.
\label{Relation_dFdG}
\eeqn
It simply yields
\beq
G(r)=F^{-1}(r),
\label{Relation_FG}
\eeq
where we have fixed the integration constant to be 1, which can be realized by rescaling the time $t$ in the metric \eqref{Metric}. 

By substituting it into the modified Maxwell equation \eqref{Maxwell_EQ_2}, we have
\beq
\Phi ''+\frac{2}{r}\Phi'=0.
\eeq
Then, the electrostatic potential is obtained as
\beq
\Phi(r)=\frac{c_1}{r}+c_2,
\eeq
where the integration constant $c_2$ can be set to zero by fixing the zero point of the potential to be zero at infinity. However, since the conserved current has been modified to be $J^{\mu}=\nabla_{\nu }\lt(F^{\mu \nu } + 2\eta  B^{\mu \nu }B^{\alpha \beta }  F_{\alpha \beta }\rt)$, the electric charge $Q$ can be determined by using Stokes's theorem \cite{Carroll2019}, i.e.,
\beqn
 Q&=& -\fc{1}{4\pi}\int_\Sigma dx^3 \sqrt{\gamma^{(3)}}n_\mu J^\mu=
 -\fc{1}{4\pi}\int_{\pt \Sigma} d\theta d\phi\sqrt{\gamma^{(2)}}n_\mu\sigma_\nu \lt(F^{\mu \nu } + 2\eta  B^{\mu \nu }B^{\alpha \beta }  F_{\alpha \beta }\rt)\nn\\
 &=&\lt(1-2 b^2 \eta \rt)c_1,
\eeqn
where $\Sigma$ represents a 3-dimensional spacelike region with the induced metric $\gamma^{(3)}_{ij}$, while its boundary $\partial\Sigma$ is a two-sphere located at spatial infinity with the induced metric $\gamma^{(2)}_{ij}=r^2 \lt (d\theta^2+\sin^2\theta d\phi^2 \rt)$. Accordingly, $n_\mu=(1,0,0,0)$ denotes the unit normal vector associated with $\Sigma$, and $\sigma_\mu=(0,1,0,0)$ denotes the unit normal vector associated with $\partial\Sigma$. Then, the integration constant $c_1$ is fixed as $c_1=Q/\lt(1-2 b^2 \eta \rt)$. Thus, the electrostatic potential is given by
\beq
\Phi(r)=\frac{Q}{\lt(1-2 b^2 \eta \rt)r}.
\eeq

After subtracting Eq.~\eqref{EoM_3} from Eq.~\eqref{EoM_1} and substituting  \eqref{Relation_FG} and \eqref{Phi} into it, one has 
\beq
\frac{F'}{F}+\frac{1+l-2 (3-l) \eta b^2}{ (1-l)^2 \left(1-2 \eta b^2  \right)^2}\fc{Q^2 }{r^3 F}-\frac{1}{(1-l) r F}+\frac{1}{r}=0.
\eeq
Then, the metric function $F(r)$ can be integrated out, resulting in
\beq
F(r)=\frac{1}{1-\ell}-\frac{2M}{r}+\fc{1+\ell-2 (3-\ell) \eta b^2  }{(1-\ell)^2\left(1-2 \eta b^2  \right)^2}\fc{Q^2}{r^2},
\eeq
where the integration constant has been determined to recover the Schwarzschild-like solution from Ref.~\cite{Yang2023a} when the electric charge $Q$ vanishes. 

Further, by substituting the obtained results into all the field equations \eqref{EoM_1}, \eqref{EoM_2}, \eqref{EoM_3}, \eqref{EOM_KR_2}, and \eqref{Maxwell_EQ_2}, it is found that the solutions are consistent only if 
\beq
\eta=\fc{\ell}{2b^2}.
\label{Realtion}
\eeq
Therefore, it is evident that in the case of Lorentz violation in spacetime, the interaction $\mathcal{L}_\tx{int}$ is indispensable to achieve a charged black hole solution. 

With the relation \eqref{Realtion}, the electrostatic potential $\Phi(r)$ and the metric function $F(r)$ can be further simplified as
\beqn
\Phi(r)&=&\frac{Q}{\lt(1-\ell \rt)r},\label{Phi}\\
F(r)&=&\frac{1}{1-\ell}-\frac{2M}{r}+\fc{Q^2}{\lt(1-\ell \rt)^2r^2}.\label{Solution_Fr_RN}
\eeqn
It is worth noting that the electrostatic potential $\Phi(r)$ has been modified due to the Lorentz-violating effect.

Consequently, the Reissner-Nordstr\"om-like (RN-like) metric is obtained as
\beqn
ds^2=-\lt(\frac{1}{1-\ell}-\frac{2M}{r}+\fc{Q^2}{\lt(1-\ell \rt)^2r^2}\rt)dt^2+\fc{dr^2}{\frac{1}{1-\ell}-\frac{2M}{r}+\fc{Q^2}{\lt(1-\ell \rt)^2r^2}}+r^2 d\theta^2+r^2 \sin^2\theta d\phi^2.\quad
\label{BH_without_CC}
\eeqn
The Lorentz-violating effect arising from the nonzero VEV of the KR field is characterized by the dimensionless parameter $\ell$, whose value is constrained to be very small based on the classical gravitational experiments within the Solar System \cite{Yang2023a}. When the Lorentz-violating parameter $\ell$ vanishes, it reduces to the standard RN metric.

\begin{figure}[t]
\begin{center}
\subfigure[~$F(r)$]  {\label{Metric_Function_RN}
\includegraphics[width=7cm,height=5cm]{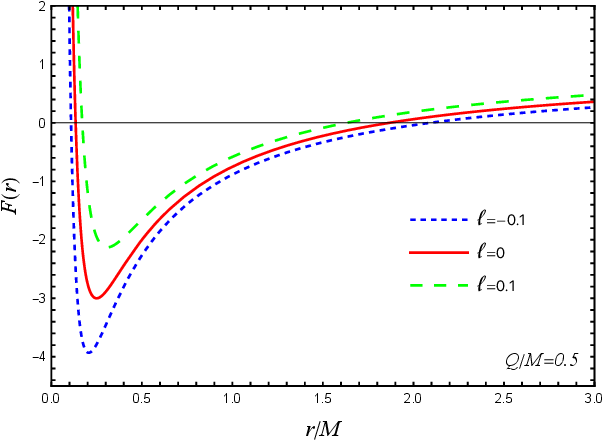}}\hfill
\subfigure[~Parameter space]  {\label{Parameter_Space_RN}
\includegraphics[width=7cm,height=5cm]{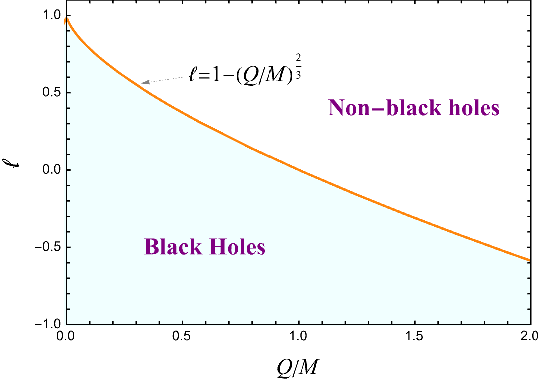}}
\end{center}
\caption{(a) Metric function $F(r)$  for the RN-like solution with varying Lorentz-violating parameter $\ell$. (b) Black hole region for the RN-dS-like solution in the parameter space $\lt(Q/M,\ell\rt)$, where the colored region represents black holes and blank region represents naked singularities.}
\end{figure}

From the metric \eqref{BH_without_CC}, the horizon radii reads  
\beq
r_\pm=(1-\ell)\lt(M \pm \sqrt{M^2-\fc{Q^2}{(1-\ell)^3}}\rt).
\label{Horizons_RN}
\eeq
When $\ell=0$, this expression recovers the result of the RN black hole. Fig.~\ref{Metric_Function_RN} illustrates that as the Lorentz-violating parameter $\ell$ increases, the outer event horizon radius $r_+$ decreases, while the inner Cauchy horizon radius $r_-$ increases. 

From the expression for the horizon radii \eqref{Horizons_RN}, it is clear that the horizons exist only when the condition ${Q^2}/{M^2}\leq (1-\ell)^3$ is satisfied, where the equality represents the case of extremal black holes. The corresponding parameter space $\lt(Q/M,\ell\rt)$ for black hole solutions and non-black hole solutions is illustrated in Fig.~\ref{Parameter_Space_RN}, where the colored region represents the black holes with horizons while the blank region represents the naked singularities. Comparing to the case of RN black holes, extremizing the charged KR black holes requires less charge for a positive $\ell$, while it requires more charge for a negative $\ell$.

Furthermore, the Kretschmann scalar for the current spacetime can be expressed as 
\beqn
R^{\alpha\beta\gamma\delta}R_{\alpha\beta\gamma\delta}=\frac{48 M^2}{r^6}\!-\!\frac{16 l M}{(1-l) r^5}\!+\!\frac{4 l^2}{(1-l)^2 r^4}+\frac{56 Q^4}{(1-l)^4 r^8}-\frac{96 M Q^2}{(1-l)^2 r^7}+\frac{8 l Q^2}{(1-l)^3 r^6}.\quad
\eeqn
Therefore, the Lorentz-violating effect cannot be eliminated solely through coordinate transformations.
It is worth noting that the metric functions approach to $F(r)=1/G(r)\to 1/(1-\ell)$ as $r\to\infty$. It can be straightforwardly verified that not all components of the Riemann tensor are zero in this case. Therefore, this indicates that the current spacetime is not asymptotically Minkowski.

\subsection{Case: $\Lambda \neq 0$}

When the cosmological constant is present,  it has been found that there is no solution that satisfies all the equations of motion under the assumption $V'(X)=0$, where $X\equiv B^{\mu\nu} B_{\mu\nu}+b^2$. Therefore, following the same approach as in Bumblebee gravity \cite{Maluf2021}, we impose a linear potential as the form $V=\lambda X$, where $\lambda$ is a Lagrange multiplier field \cite{Bluhm2008}. In this case, the vacuum condition is relaxed to be $V'(X)= \lambda$. The equation of motion of the Lagrange-multiplier $\lambda$ reads $X=0$, so the on-shell $\lambda$ guarantees that $b_{\mu\nu}$ is the vacuum configuration. As a result, the on-shell value of $\lambda$ is determined by the vacuum field equations \eqref{EOM_Exp} and \eqref{Maxwell_EQ_2}. It is noting that the off-shell $\lambda$ should have the same sign as $X$ in order to keep the potential $V$ positive  \cite{Bluhm2008}.

Despite the presence of the cosmological constant, we can still derive the same relationships from the field equations \eqref{EoM_1}, \eqref{EoM_2} and \eqref{Maxwell_EQ_2}, i.e.,
\beqn
G(r)&=&F^{-1}(r),\label{Relation_FG_2}\\
\Phi(r)&=&\frac{Q}{(1-\ell)r}.\label{Phi_2}
\eeqn
Furthermore, by subtracting Eq.~\eqref{EoM_3} from Eq.~\eqref{EoM_1} and substituting the relations \eqref{Relation_FG_2} and \eqref{Phi_2} into it, we obtain
\beqn
\frac{F'}{F}+\frac{1+l-2(3-l) \eta b^2}{(1-l)^4 }\fc{Q^2}{r^3 F}+\frac{ (1-l)\lambda b^2 +(1-3 l)\Lambda}{(1-l)^2}\fc{r}{F}
-\frac{1}{(1-l) r F}+\frac{1}{r}=0.\qquad
\eeqn
It yields
\beqn
F(r)=\frac{1}{1-\ell}-\frac{2M}{r}+\fc{1+\ell-2 (3-\ell) \eta b^2 }{(1-\ell)^4}\fc{Q^2}{r^2}-\frac{(1-3 \ell)\Lambda + (1-\ell) b^2 \lambda}{3 (1-\ell)^2}r^2.
\label{Sol_F_2}
\eeqn
Finally, by substituting Eqs.~\eqref{Relation_FG_2} and \eqref{Sol_F_2} into all the field equations \eqref{EoM_1}, \eqref{EoM_2}, \eqref{EoM_3}, \eqref{EOM_KR_2}, and \eqref{Maxwell_EQ_2}, one finds that the solutions are consistent only if
\beqn
\eta = \fc{\ell}{2b^2},\quad
\lambda = \frac{2 \ell \Lambda}{(1-\ell) b^2}.
\eeqn
It is evident that the theory supports a RN-(A)dS-like black hole solution with a non-vanishing cosmological constant if only $\eta\neq 0$ and $\lambda\neq 0$. 

\begin{figure}[t]
\begin{center}
\subfigure[~RN-AdS-like solution]  {\label{Metric_Function_RN_AdS}
\includegraphics[width=7cm,height=5cm]{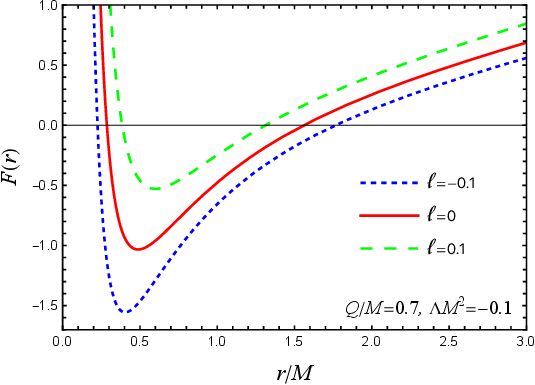}}\hfill
\subfigure[~RN-dS-like solution]  {\label{Metric_Function_RN_dS}
\includegraphics[width=7cm,height=5cm]{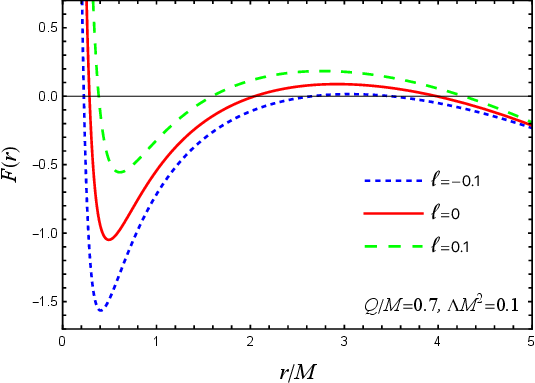}}
\end{center}
\caption{Metric function $F(r)$ for RN-AdS-like (a) and RN-dS-like (b) solutions with varying Lorentz-violating parameter $\ell$.}
\end{figure}

Consequently, the metric function $F(r)$ simplifies to
\beq
F(r)=\frac{1}{1-\ell}-\frac{2M}{r}+\fc{Q^2}{(1-\ell)^2 r^2}-\frac{\Lambda  r^2}{3 (1-\ell)}.
\label{Solution_Fr_RN_AdS}
\eeq
As a result, the RN-(A)dS-like metic is achieved as
\beqn
ds^2\!&\!=\!&\! -\!\lt(\frac{1}{1-\ell}-\frac{2M}{r}\!+\!\fc{Q^2}{(1-\ell)^2r^2} \!-\! \frac{\Lambda  r^2}{3 (1 - \ell)} \rt) dt^2+\fc{dr^2}{\frac{1}{1-\ell} \!-\! \frac{2M}{r} \!+\! \fc{Q^2}{(1-\ell)^2 r^2}\! -\! \frac{\Lambda  r^2}{3 (1-\ell)}}\nn\\
\!&\!+\!&\! r^2 d\theta^2 \!+\! r^2 \!\sin^2\!\theta d\phi^2\!.\quad\quad
\label{BH_with_CC}
\eeqn
If the cosmological constant vanishes, the solution reduces to the RN-like one \eqref{BH_without_CC}. If the electric charge vanishes, it reduces to Schwarzschild-(A)dS metric \cite{Yang2023a}. Furthermore, it degenerates to the RN-(A)dS metric when the Lorentz-violating parameter $\ell$ is set to zero.

\begin{figure}[t]
\begin{center}
\subfigure[~RN-AdS-like solution]  {\label{Parameter_Space_RN_AdS}
\includegraphics[width=7cm,height=5cm]{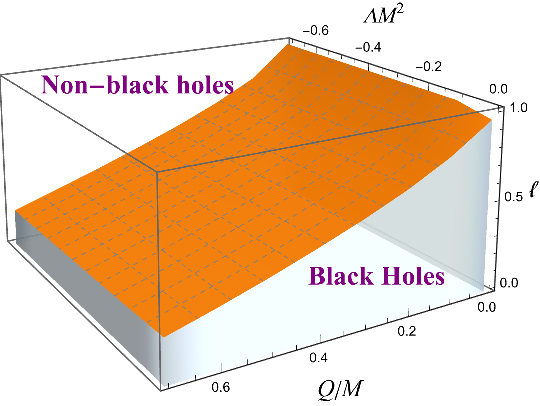}}\hfill
\subfigure[~RN-dS-like solution]  {\label{Parameter_Space_RN_dS}
\includegraphics[width=7cm,height=5cm]{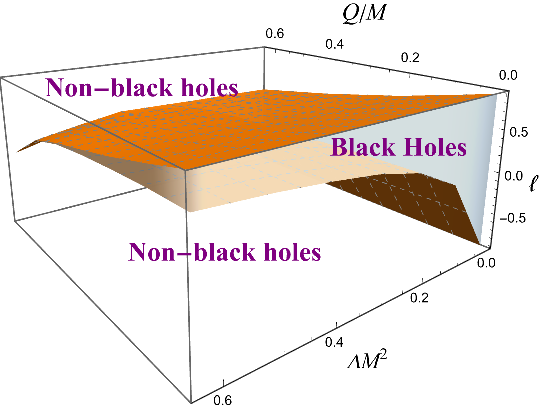}}
\end{center}
\caption{Black hole regions in parameter spaces $(Q/M,\Lambda M^2,\ell)$ for RN-AdS-like (a) and RN-dS-like (b) solutions, where the colored region represents the black hole solutions and the blank region represents the non-black hole solutions.}
\label{Parameter_Space_RN_dS&AdS}
\end{figure}

For the RN-AdS-like black hole solution, it exhibits two horizons known as the outer event horizon and the inner Cauchy horizon. Fig.~\ref{Metric_Function_RN_AdS} illustrates that as the Lorentz-violating parameter $\ell$ increases, the outer event horizon radius contracts, while the inner Cauchy horizon radius expands. When the two horizons coincide with each other, it corresponds to the formation of the extremal black hole. The corresponding parameter space $(Q/M,\Lambda M^2,\ell)$ for the RN-AdS like solutions are depicted in Fig.~\ref{Parameter_Space_RN_AdS}, where the colored region represents black holes with horizons, the orange surface represents extremal black holes, and the blank region corresponds to naked singularities.

However, as shown in Fig.~\ref{Metric_Function_RN_dS}, the RN-dS-like black hole solution exhibits three horizons, i.e., the outermost cosmological horizon, the middle event horizon, and the innermost Cauchy horizon. As the Lorentz-violating parameter $\ell$ increases, the event horizon radius contracts, while the radii of Cauchy horizon and cosmological horizon enlarge. Fig.~\ref{Parameter_Space_RN_dS} illustrates the parameter spaces $(Q/M,\Lambda M^2,\ell)$ for the RN-dS-like solutions, where the colored region corresponds to black hole solutions. Within this figure, two distinct surfaces can be observed. The upper surface represents the formation of the extremal black holes, characterized by the coincidence of the Cauchy horizon and the event horizon. On the other hand, the lower surface corresponds to the situation that the cosmological horizon and the event horizon coincide with each other.

In particular, as illustrated in Fig.~\ref{Parameter_Space_RN} and Fig.~\ref{Parameter_Space_RN_dS&AdS}, the Lorentz-violating parameter $\ell$ has a significant impact on the parameter spaces of the charged KR black hole solutions.

As $r$ approaches infinity, the metric functions approximate $F(r)=1/G(r)\to-\frac{\Lambda  r^2}{3 (1-\ell)}$. As a result of the additional contributions from the modified Einstein equations \eqref{EoM2} when $V'(X)$ is nonzero, instead of the bare cosmological constant $\Lambda$, the effective cosmological constant is given by $\Lambda_\tx{eff} \equiv \fc{\Lambda}{1-\ell}$ in this case. Similarly, due to the nontrivial contribution arising from the interaction between the electromagnetic field and the KR field \eqref{Interaction}, it was observed that the bare electric charge $Q$ is replaced by the effective charge $Q_\text{eff}=\frac{Q}{1-\ell}$ in the electrostatic potential $\Phi(r)$.

\section{Thermodynamics}\label{Thermodynamics}

One of the intriguing aspects of black holes is that they can be viewed as thermodynamic systems, governed by the laws of black hole thermodynamics \cite{Bardeen1973,Kastor2009} and exhibiting intriguing phase structures reminiscent of everyday thermodynamics \cite{Kubiznak2017, Yang2022c}. In this section, we study some basic thermodynamic properties of the charged KR black hole solutions. Due to the thermodynamics of asymptotically dS black holes is complicated and our understanding on it remains limited \cite{Mbarek2019}, we focus on the cases of the RN-like and RN-AdS-like black holes.

By solving $F(r_+)=0$ in Eq.~\eqref{Solution_Fr_RN_AdS}, the mass of the RN-AdS-like black hole can be expressed with the radius of the event horizon $r_+$, 
\beq
M=\frac{r_+}{2}\lt(\fc{1}{1-\ell}-\fc{r_+^2\Lambda_\tx{eff}}{3} \rt)+\frac{Q_\tx{eff}^2}{2 r_+}.
\label{Relation_M_rh}
\eeq
Correspondingly, the mass of the RN-like black hole is obtained by simply setting $\Lambda_\tx{eff}=0$. 

The effective cosmological constant plays the role of a thermodynamic pressure, given by \cite{Kubiznak2017}
\beq
P=-\fc{\Lambda_\tx{eff}}{8\pi}=-\fc{\Lambda}{8\pi(1-\ell)}.
\label{Pressure}
\eeq
In contrast to the RN-AdS black hole ($\ell=0$), the Lorentz-violating parameter introduces a modification to the pressure.
In this case, the black hole mass $M$ can be interpreted as a gravitational version of chemical enthalpy. Since the Lorentz-violating parameter $\ell$ is a dimensionless constant, the enthalpy can be expressed as a function of entropy $S$, pressure $P$ and effective electric charge $Q_\tx{eff}$, i.e., $M=M\lt(S,P,Q_\tx{eff}\rt)$. Therefore, the first law of the RN-AdS-like black hole is given by
\beq
dM=\mathcal{T}dS+\mathcal{V}dP+\Phi dQ_\tx{eff},
\label{First_law}
\eeq
where $\mathcal{T}$ is the Hawking temperature, $\mathcal{V}$ the thermodynamic volume, and $\Phi$ the electrostatic potential expressed as in Eq.~\eqref{Phi_2}.

By utilizing the metric \eqref{BH_with_CC} and Eq.~\eqref{Relation_M_rh}, the temperature  of the RN-AdS-like black hole is given by
\beq
\mathcal{T}\!=\!-\!\lt.\fc{1}{4\pi}\fc{\pt g_{tt}}{\pt r}\rt|_{r_+}\!=\!\frac{1}{4\pi r_+}\lt(\fc{1}{1\!-\!\ell} \!-\!r_+^2\Lambda_\tx{eff}\rt)\!-\!\frac{Q_\tx{eff}^2}{4 \pi r_+^3}.
\label{Temperature}
\eeq
The result of the RN-like black hole is obtained by directly setting $\Lambda_\tx{eff}=0$. 

From the first law \eqref{First_law} and the Hawking temperature \eqref{Temperature}, the entropy can be shown to satisfy the standard Bekenstein-Hawking area-entropy relation, i.e.,
\beq
S\!=\!\int{}\!\lt(\!\fc{d M}{\mathcal{T}} \!\rt)_{P,Q_\tx{eff}}\!=\!\int{}\!\fc{1}{\mathcal{T}}\!\lt(\!\fc{\pt M}{\pt r_+}\! \rt)_{P,Q_\tx{eff}} \!d r_+\!=\!\fc{A_+}{4},
\eeq
where $A_\tx{+}=4\pi r_\tx{+}^2$ is the area of the event horizon.

Moreover, from the first law \eqref{First_law} and the pressure \eqref{Pressure}, we can also calculate the thermodynamic volume, given by 
\beq
\mathcal{V}\!=\!\lt(\fc{\pt M}{\pt P}\rt)_{S,Q_\tx{eff}}\!=\!\lt(\fc{\pt M}{\pt \Lambda }\rt)_{S,Q_\tx{eff}}\!\lt(\fc{\pt \Lambda}{\pt P }\rt)_{S,Q_\tx{eff}}\!=\!\frac{4 \pi r_+^3}{3}.
\eeq

Now, with the aforementioned results, it can be straightforwardly verified that the Smarr formula holds in the same form as the RN-AdS black hole, i.e.,
\beq
M=2\mathcal{T}S-2\mathcal{V}P+\Phi Q_\tx{eff}.
\eeq 

\begin{figure}[t]
\begin{center}
\subfigure[~RN-AdS-like black hole]  {\label{Plot_Heat_capacity_AdS}
\includegraphics[width=7cm,height=5cm]{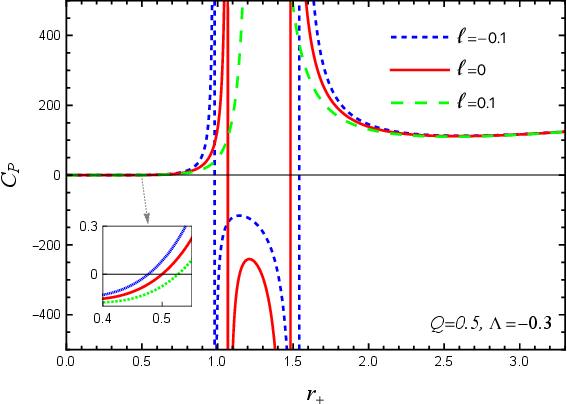}}\hfill
\subfigure[~RN-like black hole]  {\label{Plot_Heat_capacity}
\includegraphics[width=7cm,height=5cm]{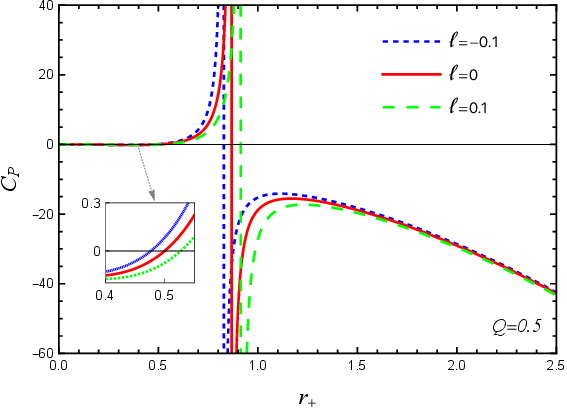}}
\end{center}
\caption{Heat capacity $C_P$ versus horizon radius $r_+$ for RN-AdS-like (a) and RN-like (b) black holes with varying Lorentz-violating parameter $\ell$.}
\label{Heat_capacity}
\end{figure}

In order to analyze the thermodynamical stabilities of the charged KR black holes, we can evaluate their heat capacity, where a positive heat capacity indicates the local stability of the black hole. The heat capacity of the RN-AdS-like black holes is determined by
\beqn
C_P=T \lt(\fc{\pt S}{\pt \mathcal{T}}\rt)_{P}=T\lt(\fc{\pt S}{\pt r_+ }\rt)_{P}\lt(\fc{\pt r_+}{\pt \mathcal{T} }\rt)_{P}=2 \pi  r_+^2 \frac{\lt(r_+^2\Lambda_\tx{eff}-\fc{1}{1-\ell} \rt)r_+^2+Q_\tx{eff}^2}{\lt(r_+^2\Lambda_\tx{eff}+\fc{1}{1-\ell}\rt)r_+^2-3Q_\tx{eff}^2}.
\label{Heat_capacity_RN_Ads}
\eeqn
It is found that that for $l<1+ 12 \Lambda  Q^2$, the heat capacity is positive within the intervals $r_+ \subset \Bigg(\sqrt{\frac{\sqrt{1-4(1-\ell)^2\Lambda_\tx{eff} Q_\tx{eff}^2}-1}{-2(1-\ell)\Lambda_\tx{eff} }},\sqrt{\frac{1-\sqrt{1+12(1-\ell)^2\Lambda_\tx{eff}  Q_\tx{eff}^2}}{-2 (1-\ell)\Lambda_\tx{eff}}}\Bigg)\cup \Bigg(\sqrt{\frac{1+\sqrt{1+12(1-\ell)^2\Lambda_\tx{eff}  Q_\tx{eff}^2}}{-2 (1-\ell)\Lambda_\tx{eff}}},\infty \Bigg)$, while for $l\geq 1+ 12 \Lambda  Q^2$, the heat capacity is positive for  
$r_+\subset\lt(\sqrt{\frac{\sqrt{1-4(1-\ell)^2\Lambda_\tx{eff} Q_\tx{eff}^2}-1}{-2(1-\ell)\Lambda_\tx{eff} }},\infty\rt)$.
As depicted in Fig.~\ref{Plot_Heat_capacity_AdS}, an increase in the Lorentz-violating parameter $\ell$ leads to an expansion of the ranges that guarantees local stability. In comparison to the RN-AdS black holes, the class of RN-AdS-like black holes with $\ell>0$ demonstrates a greater stable region, while for $\ell<0$, it exhibits a smaller stable region.

For the RN-like black holes, the heat capacity is obtained by setting $\Lambda_\tx{eff}=0$ in Eq.~\eqref{Heat_capacity_RN_Ads}, yielding
\beq
C_P=-2 \pi  r_+^2 \frac{r_+^2-(1-\ell)Q_\tx{eff}^2}{r_+^2-3(1-\ell)Q_\tx{eff}^2}.
\eeq
The heat capacity is positive only for the horizon radii within the range $r_+\subset\big(\sqrt{1-\ell}Q_\tx{eff},$ $\sqrt{3(1-\ell)}Q_\tx{eff}\big)$, just as shown in Fig.~\ref{Plot_Heat_capacity}.  When compared to the RN black hole, the stable region of the RN-like black holes with $\ell>0$ is larger, while for $\ell<0$, it is smaller.

Interestingly, if the electric charge is set to zero, the heat capacity of the RN-AdS-like and RN-like black holes is independent of the Lorentz-violating parameter $\ell$.

\begin{figure}[t]
\begin{center}
\subfigure[~Different $P$]  {\label{Free_Energy_vs_temperature}
\includegraphics[width=7cm,height=5cm]{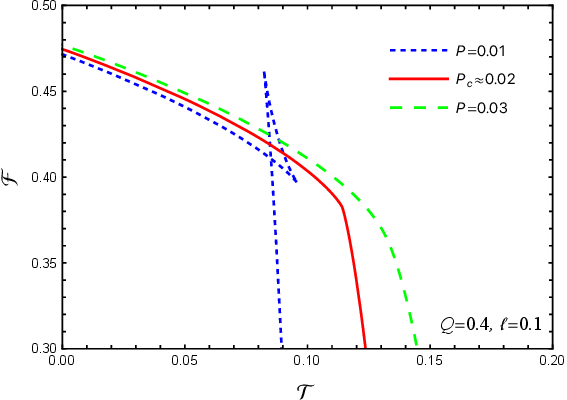}}\hfill
\subfigure[~Different $\ell$]  {\label{Free_Energy_vs_l}
\includegraphics[width=7cm,height=5cm]{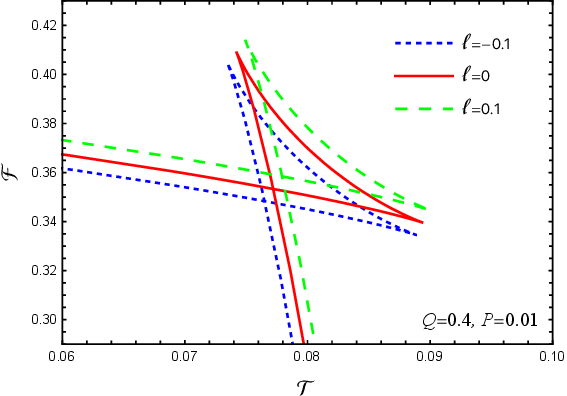}}
\end{center}
\caption{Free energy $\mathcal{F}$ versus temperature $\mathcal{T}$ for RN-AdS-like black holes with varying pressure $P$ (a) and Lorentz-violating parameter $\ell$ (b).}
\label{Free_energy}
\end{figure}

Furthermore, it is well known that the charged AdS black holes allow for a first-order phase transition between small black holes and large black holes in the canonical ensemble \cite{Kubiznak2017}. For the RN-AdS-like black holes, the Gibbs free energy in the canonical ensemble can be calculated as
\beq
\mathcal{F}=M-\mathcal{T}S=\frac{r_+}{4}\lt(\fc{1}{1-\ell}+\fc{r_+^2\Lambda_\tx{eff}}{3}\rt)+\frac{3 Q_\tx{eff}^2}{4 r_+}.
\eeq
The $\mathcal{F}-\mathcal{T}$ diagram is plotted in Fig.~\ref{Free_energy}. As depicted in Fig.~\ref{Free_Energy_vs_temperature}, the swallowtail behavior emerges for pressures below the critical pressure $P_\text{c}$. This swallowtail behavior signifies a first-order phase transition. At the critical pressure $P_\tx{c}$, the phase transition becomes second-order, while above the critical pressure $P_\tx{c}$, no phase transition occurs. The critical point of second-order phase transition can be analytically determined as
\beqn
P_\tx{c}&=&\frac{1}{96 \pi (1-\ell)^2  Q_\tx{eff}^2}=\frac{1}{96  \pi  Q^2}, \\
\mathcal{T}_\tx{c}&=&\frac{1}{3 \sqrt{6}\pi (1-\ell)^\fc{3}{2} Q_\tx{eff}}=\frac{1}{3\pi  \sqrt{6(1-\ell)} Q},\\
r_\tx{c+}&=&\sqrt{6(1- \ell)} Q_\tx{eff}=\sqrt{\fc{6}{1- \ell}} Q.
\eeqn
It is evident that the critical pressure $P_\tx{c}$ is independent of the Lorentz-violating parameter $\ell$.
So it exhibits the same critical pressure as the RN-AdS black holes. However, both the critical temperature $\mathcal{T}_\tx{c}$ and the critical size $r_\tx{c+}$ of the black holes are influenced by the Lorentz-violating effect. Therefore, in comparison to the RN-AdS black holes, the RN-AdS-like black holes with $\ell>0$ exhibit a higher critical temperature $\mathcal{T}_\tx{c}$ and a larger critical size $r_\tx{c+}$, whereas the RN-AdS-like black holes with $\ell<0$ display a lower critical temperature $\mathcal{T}_\tx{c}$ and a smaller critical size $r_\tx{c+}$.

The intersection point of the swallowtail represents the first-order phase transition point, corresponding to the coexistence phase of a large black hole and a small black hole. As illustrated in Fig.~\ref{Free_Energy_vs_l}, it can be observed that, for a fixed charge and pressure below the critical pressure $P_\tx{c}$, the first-order phase transition temperature increases with the Lorentz-breaking parameter $\ell$. 

\section{Orbital motion of test particles}\label{Orbits}

In this section, our focus is on studying the effects induced by Lorentz violation on the motion of massless photons and massive particles in the vicinity of the charged KR black holes. 

The motion of a test particle along its geodesics is governed by the Lagrangian
\beq
\mathcal{L}=-\fc{1}{2}g_{\alpha\beta}\dot{x}^\alpha\dot{x}^\beta,
\label{Lagrangian_Abst}
\eeq
where the dot represents the derivative with respect to the affine parameter $\lambda$. From the normalization conditions of 4-velocity, one can express the Lagrangian density as $\mathcal{L}=\epsilon/2$, with $\epsilon=1$ for massive particles and $\epsilon=0$ for photons. 

Since the spacetime is spherically symmetric, without loss of generality, we restrict ourselves to the study of equatorial geodesics with $\theta=\pi/2$. Consequently, the Lagrangian density \eqref{Lagrangian_Abst} leads to the equation
\beq
F(r) \lt(\fc{dt}{d\lambda}\rt)^{2} - F(r)^{-1} \lt(\fc{dr}{d\lambda}\rt)^{2}-r^{2} \lt(\fc{d\phi}{d\lambda}\rt)^{2}=\epsilon.
\label{Lagrangian_Expl}
\eeq
Further, by taking into account two conserved quantities of the static and spherically symmetric spacetime, i.e., the energy per unit mass $E= \fc{\pt \mathcal{L}}{\pt \dot{t} }=F(r)\fc{dt}{d\lambda}$ and the angular momentum per unit mass $L =-\fc{\pt \mathcal{L}}{\pt \dot{\phi}}=r^{2}\fc{d\phi }{d \lambda}$, we can derive 
\beq
\dot{r}^2=E^2-F(r) \lt(\epsilon^2+\fc{L^2}{r^2}\rt)=E^2-V^2_\tx{eff},
\label{Orbital_Eq}
\eeq
with the effective potential defined by $V_\tx{eff}=\sqrt{F(r) \lt(\epsilon^2+\fc{L^2}{r^2}\rt)}$.
Therefore, through an analysis of the effective potential, we can gain insights into the motion of particles in the vicinity of black holes. 

\subsection{Shadow of massless photons}

For the massless photons, it is interesting to study the formation of a black hole shadow, 
which is a result of the interaction between the intense gravitational field of the black hole and the surrounding light rays. When the light rays  pass near to a black hole, they are deflected very strongly by its gravitational field.  The photons with small orbital angular momentum are trapped by the black hole, and only those photons with large orbital angular momentum can escape from it. As a result, a distant observer observes a dark zone in the sky, known as the black hole shadow. 

The observations of black hole shadows of M87* and Sagittarius A* by the Event Horizon Telescope (EHT) provide an unprecedented window onto tests of gravitational theory and fundamental physics in the strong-field regime \cite{Akiyama2019,Akiyama2022a}. Consequently, investigating the influence of Lorentz violation on black hole shadows allows us to utilize experimental data to constrain the magnitude of the Lorentz violation.

For the photon, where $\epsilon=0$, we can rewrite Eq.~\eqref{Orbital_Eq} as
\beq
\dot{r}^2=E^2-\fc{L^2}{r^2}F(r)=E^2-V^2_\tx{eff},
\eeq
with the effective potential given by $V_\tx{eff}={L\sqrt{F(r)}}/{r}$.

For the circular photon orbit, the effective potential satisfies the conditions \cite{Guo2020}
\beq
V_\tx{eff}=E, \quad \fc{\pt V_\tx{eff}}{\pt r}=0, \quad \tx{and} ~~ \fc{\pt^2 V_\tx{eff}}{\pt r^2}<0.
\eeq
Consequently, the radius of the circular photon orbit is determined by the implicit equation
\beq
r_\tx{ph}=2\fc{F(r_\tx{ph})}{F'(r_\tx{ph})}.
\eeq
Due to the inherent spherical symmetry, the photons can occupy all circular orbits, resulting in the formation of the photon sphere. Using the metric function \eqref{Solution_Fr_RN_AdS}, the radius $r_\tx{ph}$ of the photon sphere can be obtained as
\beq
r_\tx{ph}=\frac{3(1-\ell) M}{2}\lt[1+ \sqrt{1-\fc{8 Q^2}{9 (1-\ell)^3 M^2}}\rt].
\label{Photon_sphere_radius_RN_AdS}
\eeq
When $\ell\to 0$, it recovers the result of the RN-(A)dS black hole \cite{Eiroa2002a}, i.e.,
\beq
r_\tx{ph}=\fc{3M}{2}\lt(1+\sqrt{1-\fc{8Q^2}{9M^2}}\rt).
\eeq

\begin{figure}[t]
\begin{center}
\subfigure[~RN-AdS-like black hole]  {\label{Shadow_RN_AdS}
\includegraphics[width=7cm,height=7cm]{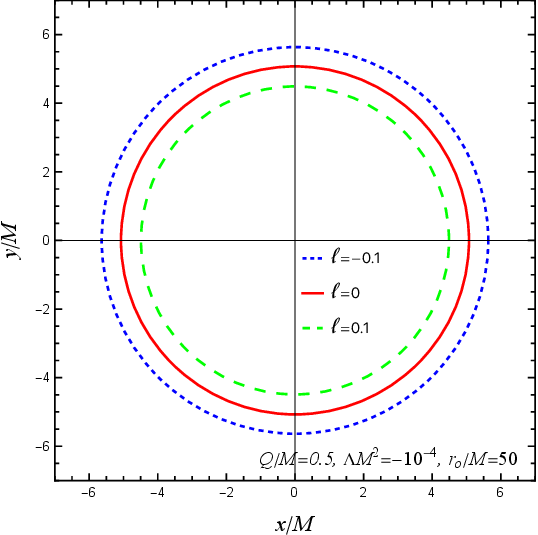}}\hfill
\subfigure[~RN-dS-like black hole]  {\label{Shadow_RN_dS}
\includegraphics[width=7cm,height=7cm]{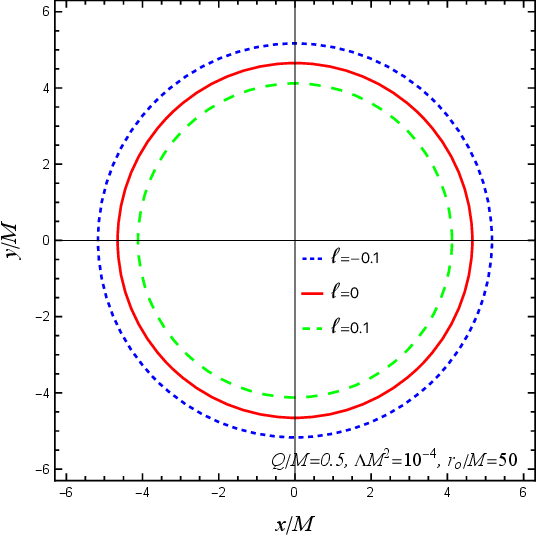}}
\end{center}
\caption{Shadows cast by RN-AdS-like (a) and RN-dS-like (b) black holes observed in the celestial coordinates with varying Lorentz-violating parameter $\ell$.}
\label{Shadow}
\end{figure}

The shadow radius of the RN-(A)dS-like black-hole observed by a static observer at the position $r_\tx{o}$ is given by \cite{Zhang2020f}
\beq
r_\tx{sh}\!=\!\!\sqrt{\fc{F(r_\tx{o})}{F(r_\tx{ph})}}r_\tx{ph}\!=\!\!\sqrt{\! \fc{\frac{1}{1-\ell}\!-\!\frac{2M}{r_\tx{o}}\!+\!\fc{Q^2}{(1-\ell)^2r_\tx{o}^2}-\frac{\Lambda  r_\tx{o}^2}{3 (1-\ell)}}{\!\frac{1}{1-\ell}\!-\!\frac{2M}{r_\tx{ph}}\!+\!\fc{Q^2}{(1-\ell)^2r_\tx{ph}^2}\!-\!\frac{\Lambda  r_\tx{ph}^2}{3 (1-\ell)} }}r_\tx{ph}.
\label{Shadow_radius_RN_AdS}
\eeq
By substituting the photon sphere radius \eqref{Photon_sphere_radius_RN_AdS} into this equation, we can obtain an explicit formula for the shadow radius of the RN-(A)dS-like black hole. Since the exact expression is somewhat cumbersome, we present some numerical results to illustrate the findings, as shown in Fig.~\ref{Shadow}.

For the RN-like black hole, $F(r_\tx{o})$ approaches $\frac{1}{1-\ell}$ for the observer located at infinity. In this case, the formula \eqref{Shadow_radius_RN_AdS} simplifies to
\beq
r_\tx{sh}=\frac{3 \sqrt{3} (1-\ell) M\left(1+\sqrt{1-\frac{8 Q^2}{9 (1-\ell)^3 M^2}}\right)^2}{2 \sqrt{2}\sqrt{1-\frac{2 Q^2}{3 (1-\ell)^3 M^2}+\sqrt{1-\frac{8 Q^2}{9 (1-\ell)^3 M^2}}}}.
\label{Shadow_radius_RN}
\eeq
If we further set the charge $Q$ to zero, we obtain the shadow radius of the Schwarzschild-like black hole, i.e.,
\beq
r_\tx{sh}=3 \sqrt{3} (1-\ell) M.
\label{Shadow_radius_Schwarzschild}
\eeq 

From Eqs.~\eqref{Shadow_radius_RN_AdS}, \eqref{Shadow_radius_RN}, and \eqref{Shadow_radius_Schwarzschild}, it is evident that the Lorentz-violating parameter deforms the sizes of the black hole shadows. As shown in Fig.~\ref{Shadow}, the shadow size of these black holes decreases with the Lorentz-violating parameter $\ell$.

\begin{figure}[t]
\begin{center}
\subfigure[~RN-AdS-like black hole]  {\label{Shadow_VS_All_RN_AdS}
\includegraphics[width=7cm,height=5cm]{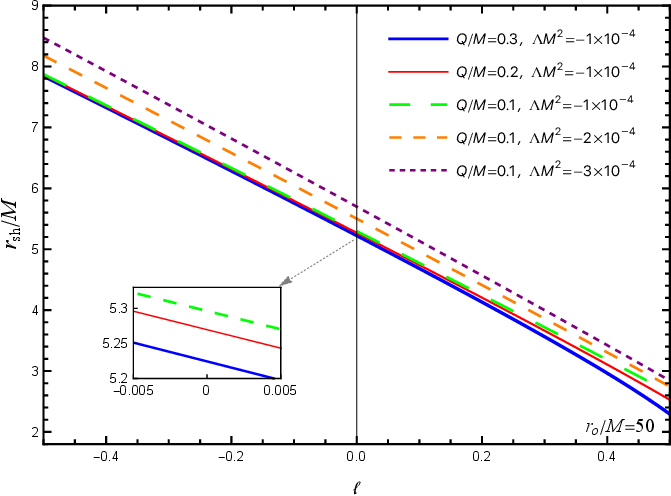}}\hfill
\subfigure[~RN-dS-like black hole]  {\label{Shadow_VS_All_RN_dS}
\includegraphics[width=7cm,height=5cm]{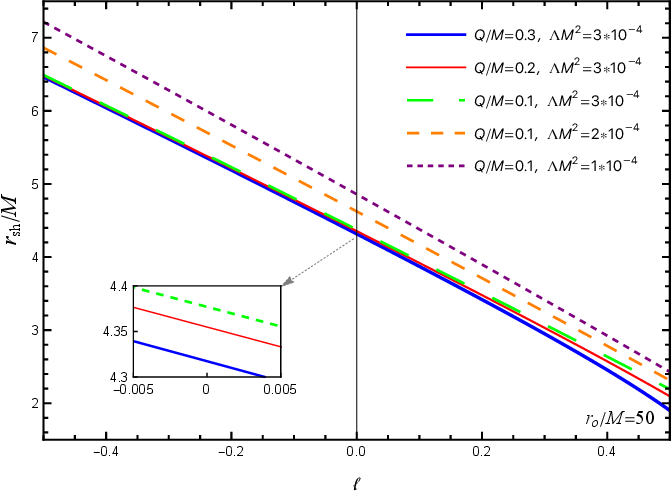}}
\end{center}
\caption{Shadows radius $r_\tx{sh}/M$ for RN-AdS-like (a) and RN-dS-like (b) black holes with varying the Lorentz-violating parameter $\ell$, electric charge $Q/M$, and cosmological constant $\Lambda M^2$.}
\label{Shadow_VS_All}
\end{figure}

The dependance of the shadow radius on additional parameters $Q$ and $\Lambda$ is illustrated in Fig.~\ref{Shadow_VS_All}. It is evident that as the electric charge $Q$ increases, both the RN-AdS-like and RN-dS-like black holes exhibit a reduction in their shadow radii. Moreover, when the magnitude of the cosmological constant $\Lambda$ increases, the shadow radius of the RN-AdS-like black hole increases, whereas the shadow radius of the RN-dS-like black hole decreases. Especially, it reveals that the shadow radius is more sensitive to variations in the Lorentz-violating parameter $\ell$ and the cosmological constant $\Lambda$ than to the electric charge $Q$.

Furthermore, the measurements of the shadow radius, determined by the average between the estimates based on the VLTI and Keck instruments within a $1\sigma$ uncertainty, result in the following restriction \cite{Vagnozzi2023}:
\beq
4.55 M \leq r_\tx{sh} \leq 5.22 M.
\label{Constraint_1s}
\eeq
Hence, with the restriction, we can derive the constraint for the Lorentz-violating parameter $\ell$ from Eq.~\eqref{Shadow_radius_Schwarzschild} as $-4.59 \times 10^{-3} \leq \ell \leq 1.24 \times 10^{-1}$. This constraint is much weaker than the constraint of $- 6.1 \times 10^{-13} \leq \ell \leq 2.8 \times 10^{-14}$ obtained from the test within the Shapiro time-delay effect \cite{Yang2023a}.

When $\ell=0$, the shadow radius \eqref{Shadow_radius_RN} reduces to the result of the RN-(A)dS black hole \cite{Perlick2022}, i.e., 
\beq
r_\tx{sh}=\frac{\left(3 M+\sqrt{9 M^2-8 Q^2}\right)^2}{2 \sqrt{2} \sqrt{3 M^2-2 Q^2+M \sqrt{9 M^2-8 Q^2}}}.
\eeq
If we further set the charge $Q$ to zero, the shadow radius will further degenerate into the result for the Schwarzschild black hole, i.e., $r_\tx{sh}=3\sqrt{3}M$.

\subsection{ISCO of massive particles}

The motion of massive particles in the vicinity of black holes provides an effective description for the extreme-mass-ratio-inspiral (EMRI) system. Among the particle trajectories, there exist a special type known as the stable circular orbits. These orbits correspond to the scenario where particles rotate along the local minimum of the effective potential well, satisfying the conditions $V_\tx{eff}=E$, $\fc{dV_\tx{eff}}{dr}=0$, and $\fc{d^2V_\tx{eff}}{dr^2}>0$ \cite{Yang2022}. The minimum radius of these circular orbits, corresponding to the critical case $\fc{d^2V_\tx{eff}}{dr^2}=0$, is the so-called ISCO. In the Novikov-Thorne accretion disk model, the ISCO marks the inner edge of the black hole accretion disk. Moreover, in the context of quasi-circular, inspiraling compact binaries, the ISCO represents the point where the nature of the orbit, and consequently, the characteristics of the gravitational waves emitted, undergoes an abrupt change. Therefore, the impact of Lorentz violation on the ISCO may be examined in future experimental tests in realistic astrophysics.

For massive particles, where $\epsilon=1$, Eq.~\eqref{Orbital_Eq} becomes
\beq
\dot{r}^2=E^2-F(r) \lt(1+\fc{L^2}{r^2}\rt)=E^2-V^2_\tx{eff},
\label{Orbital_Eq_Particles}
\eeq
with the effective potential given by $V_\tx{eff}=\sqrt{F(r) \lt(1+\fc{L^2}{r^2}\rt)}$. 

From the conditions $V_\tx{eff}=E$ and $\fc{dV_\tx{eff}}{dr}=0$, we have
\beqn
E^2\!&\!=\!&\!\frac{2 F(r)^2}{2 F(r)\!-\!r F'(r)}\!=\!\frac{\left(\frac{1}{1-\ell}\!-\!\frac{2 M}{r}\!+\!\frac{Q^2}{(1-\ell)^2r^2}\!-\!\frac{\Lambda  r^2}{3 (1-\ell)}\right)^2}{\frac{1}{1-\ell}-\frac{3 M}{r}+\frac{2 Q^2}{(1-\ell)^2r^2}},\label{Condition_E2}\\
L^2\!&\!=\!&\!\frac{r^3 F'(r)}{2 F(r)-r F'(r)}\!=\!\frac{\left(\frac{M}{r}-\frac{Q^2}{(1-\ell)^2r^2}-\frac{\Lambda  r^2}{3 (1-\ell)}\right)r^2 }{\frac{1}{1-\ell}-\frac{3 M}{r}+\frac{2 Q^2}{(1-\ell)^2r^2}}.\label{Condition_L2}
\eeqn 
Consequently, with the critical condition $\fc{d^2V_\tx{eff}}{dr^2}=0$, and Eqs.~\eqref{Condition_E2} and \eqref{Condition_L2}, the radius of ISCO can be solved from the equation,
\beq
r_\tx{ISCO} = \frac{3 F(r_\tx{ISCO} ) F'(r_\tx{ISCO} )}{2 F'(r_\tx{ISCO} )^2-F(r_\tx{ISCO} ) F''(r_\tx{ISCO} )}.
\eeq

\begin{figure}[t]
\begin{center}
\includegraphics[width=7cm,height=5cm]{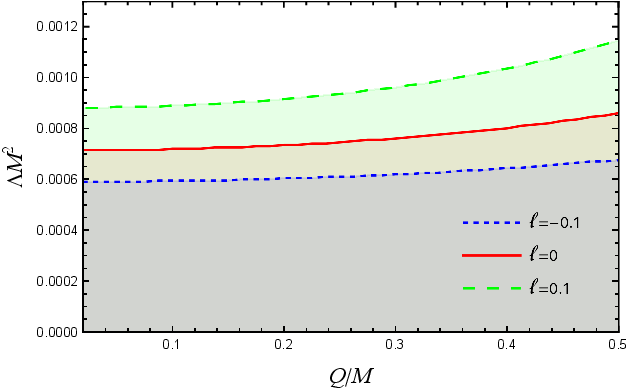}
\end{center}
\caption{Parameter space of the existence of ISCO for the RN-dS-like black hole  with varying the Lorentz-violating parameter $\ell$.} 
\label{Prameter_Space_ISCO_RN_dS}
\end{figure}

With the metric function $F(r)$ of the RN-(A)dS-like black hole \eqref{Solution_Fr_RN_AdS}, it yields 
\beqn
\frac{12 Q^4}{(1-\ell)^4 r_\tx{ISCO}^4}-\left(\frac{9 M}{r_\tx{ISCO}}-\frac{4 \Lambda  r_\tx{ISCO}^2}{1-\ell}\right)\frac{3 Q^2}{(1-\ell)^2 r_\tx{ISCO}^2}
-
\left(\!\!\frac{1}{1\!-\!\ell}\!-\!\frac{6 M}{r_\tx{ISCO}}\right)\!\frac{3 M }{r_\tx{ISCO}}\qquad\nn\\
\!+\!\left(\!\frac{4}{1\!-\!\ell}\!-\!\frac{15 M}{r_\tx{ISCO}}\right)\!\frac{\Lambda  r_\tx{ISCO}^2}{1-\ell}\!=\!0.
\label{ESCO_Eq}
\eeqn
It is evident that the first term of the equation dominates for small $r_\text{ISCO}$ and it is positive. For the RN-like black hole, the dominant term for large $r_\text{ISCO}$ is $-\frac{3M}{(1-\ell)r_\text{ISCO}}$, which is negative. Thus, the equation always possesses roots in this case. For the RN-(A)dS-like black hole, the dominant term is $\frac{4\Lambda  r_\tx{ISCO}^2}{(1-\ell)^2}$, which depends solely on the sign of $\Lambda$. As a result, the equation always possesses roots when the cosmological constant is negative. However, a root may not exist when the  cosmological constant is positive. Therefore, we numerically solved the the parameter space of the existence of ISCO for the RN-dS-like black hole, and the results for different Lorentz-violating parameter $\ell$ are presented in Fig.~\ref{Prameter_Space_ISCO_RN_dS}. It reveals that the parameter space expands with an increase in the Lorentz-violating parameter $\ell$.

\begin{figure}[t]
\begin{center}
\includegraphics[width=7cm,height=5cm]{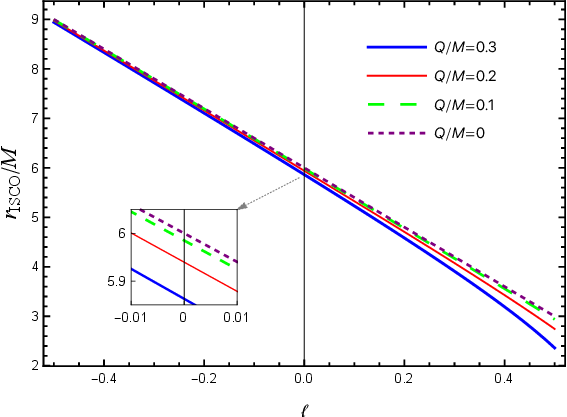}
\end{center}
\caption{ISCO radius for the RN-like black hole with varying the Lorentz-violating parameter $\ell$ and electric charge $Q/M$.}
\label{ISCO_Radius_RN}
\end{figure}

For the RN-like black hole \eqref{BH_without_CC},  Eq.~\eqref{ESCO_Eq} can be solved analytically, yielding
\beq
r_\tx{ISCO}=2(1-\ell)M\fc{1-\fc{3Q^2}{4(1-\ell)^3 M^2}+\Xi^\fc{1}{3}+\Xi^\fc{2}{3}}{\Xi^\fc{1}{3}},
\eeq
where 
\beqn
\Xi\equiv 1-\frac{9 Q^2}{8 (1-\ell)^3 M^2}+\frac{Q^4}{4 (1-\ell)^6 M^4}+\frac{Q^2 }{8 (1-\ell)^3 M^2}\sqrt{5-\frac{9 Q^2}{(1-\ell)^3 M^2}+\frac{4 Q^4}{(1-\ell)^6 M^4}}.\quad
\eeqn
If we further set the electric charge $Q$ to be zero, the ISCO radius of the Schwarzschild-like black hole is given by 
\beq
r_\tx{ISCO}=6(1-\ell)M.
\eeq
As shown in Fig.~\ref{ISCO_Radius_RN}, the ISCO radius of the RN-like black hole decreases with both the Lorentz-violating parameter $\ell$ and the electric charge $Q$.

\begin{figure}[t]
\begin{center}
\subfigure[~RN-AdS-like black hole]  {\label{ISCO_VS_l_RN_AdS}
\includegraphics[width=7cm,height=5cm]{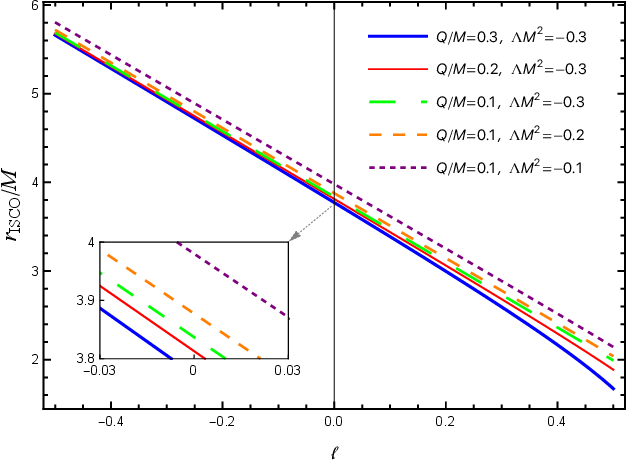}}\hfill
\subfigure[~RN-dS-like black hole]  {\label{ISCO_VS_l_RN_dS}
\includegraphics[width=7cm,height=5cm]{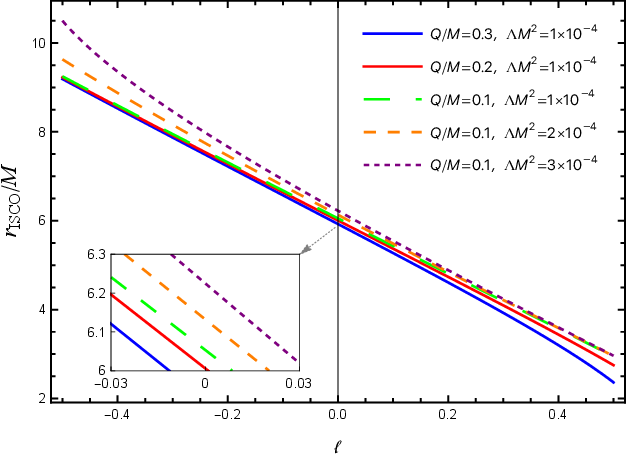}}
\end{center}
\caption{ISCO radius for the RN-(A)dS-like black hole for different parameters with varying the Lorentz-violating parameter $\ell$, electric charge $Q/M$, and cosmological constant $\Lambda M^2$.}
\label{ISCO_Radius_RN_(A)dS}
\end{figure}

However, for the RN-(A)dS-like black hole, the ISCO radius cannot be determined analytically. Instead, we present some numerical results in Fig.~\ref{ISCO_Radius_RN_(A)dS}. These results indicate that the ISCO radius of the RN-(A)dS-like black hole shrinks with an increase in the Lorentz-violating parameter $\ell$ and the charge $Q$, similar to the behavior observed in the case of the RN-like black hole. Moreover, Fig.~\ref{ISCO_VS_l_RN_AdS} shows that the ISCO radius of the RN-AdS-like black hole decreases as the magnitude of the cosmological constant increases. However, Fig.~\ref{ISCO_VS_l_RN_dS} indicates that the ISCO radius of the RN-dS-like black hole expands with an increase in the magnitude of the cosmological constant.

It is worth noting that the results presented Figs.~\ref{ISCO_Radius_RN} and \ref{ISCO_Radius_RN_(A)dS} reveal that the ISCO radius exhibits a higher sensitivity to variations in the Lorentz-violating parameter $\ell$ compared to the cosmological constant $\Lambda$ and electric charge $Q$.

\section{Conclusions}\label{Conclusions}

In this work, we derived exact RN-like and RN-(A)dS-like black hole solutions in the presence of a nonzero VEV background of the KR field. In the absence of the Lorentz violation ($\ell=0$), these solutions reduce to the RN and RN-(A)dS metrics, respectively. It was found that the Lorentz-violating parameter $\ell$ has a significant impact on the parameter spaces of black hole solutions. For instance, in contrast to the RN black hole, extremizing the charged KR black holes requires a smaller amount of electric charge when $\ell$ is positive, while it necessitates a greater amount of electric charge when $\ell$ is negative.

The thermodynamic properties of the RN-like and RN-AdS-like black holes were investigated. Our findings revealed that the standard first law of thermodynamics and the Smarr formula still hold, except for replacing the bare cosmological constant $\Lambda$ with the effective cosmological constant $\Lambda_\tx{eff}$ and the bare charge $Q$ with the effective charge $Q_\tx{eff}$. The locally stable ranges of the RN-like and RN-AdS-like black holes increase with the Lorentz-violating parameter $\ell$. Furthermore, the phase transition of the RN-AdS-like black holes was investigated. It was found that the first-order phase transition temperature increases with the Lorentz-breaking parameter $\ell$. Additionally, for the second-order phase transition, the critical pressure $P_\tx{c}$ is independent of the parameter $\ell$. However, both the critical temperature $\mathcal{T}_\tx{c}$ and the critical size $r_\tx{c+}$ increase with the parameter $\ell$.

Furthermore, in order to explore the impact of Lorentz violation on the motion of test particles near the black holes, we studied the shadow of the massless photons and ISCO of the massive particles for the charged KR black holes. Our results demonstrate that the shadow and ISCO radii of these black holes decrease with the Lorentz-violating parameter $\ell$. In particular, the radii exhibit a high sensitivity to variations in $\ell$.

In summary, our study presents exact charged black hole solutions in the Lorentz-violating gravity with a background KR field, and unveils the impact of Lorentz-violating effect on the black hole thermodynamics, shadows, and ISCO. These findings contribute to our enhanced understanding of the unique characteristics exhibited by black holes in the Lorentz-violating gravity.

It is well known that the static spherically symmetric solutions provide a good approximation for the exterior spacetime of celestial bodies in many cases, such as the classical tests in the Solar System. However, it is universally recognized that all celestial bodies rotate on their axes. Therefore, the study of rotating solutions in this gravity is an important issue, which will be left for our further investigation.

\section*{ACKNOWLEDGMENTS}

We thank Si-Jiang Yang, Yun-Zhi Du, and Wen-Di Guo for helpful discussions. This work was supported by the National Natural Science Foundation of China (Grant No.~12005174) and the Natural Science Foundation of Chongqing (Grant No.~2024NSCQ-MSX3796).


\end{document}